\documentclass[twocolumn,twoside,slac_two]{revtex4}
\usepackage{graphicx}
\usepackage{fancyhdr}
\begin{document}

\title{Using Simulation, Comparison and Grid Search 
To Find all Possible Binary Black Hole Source Parameters 
For Extreme Mass-Ratio Inspiral Gravitational Wave Signals.}

%

\author{James S. Graber}
\affiliation{Graber Research, Washington, DC 20003, USA}

\begin{abstract}
First, for each case to be tested, a specific target inspiral signal is selected for parameter extraction.  In a future real analysis, the target signal would be a real signal actually observed by a gravitational wave detector such as LISA.  In this study, however, the target signals are themselves simulations.  Some cases were selected to resemble sources likely to be detected by LISA when it flies; others were selected to facilitate comparison with previous work using Fisher matrix techniques [e.g. Leor Barack, Curt Cutler, Phys.Rev. D69 (2004) 082005].

Then, for each target inspiral signal, a grid search of the input parameter space is conducted to determine the set of input parameters that produce a simulated inspiral output signal compatible with the target.  In this study, we consider four parameters:  the two masses, the spin of the larger black hole, and the eccentricity of the orbit.  Searching through this four-dimensional parameter space requires that hundreds of possible input source parameter combinations be simulated for each target signal analyzed.  For each input parameter combination, the detailed time history of the phase of the resulting inspiral is simulated and directly compared with the phase history of the target signal. 

The simulation, comparison, and grid search technique used in this study requires more work than the Fisher matrix technique used in most previous studies of this topic.  However, this method yields a detailed map of the acceptable region of input parameter space, in contrast to the multidimensional ellipsoids of the Fisher matrix method.  Nevertheless, the final results are in general agreement with those obtained previously by the Fisher matrix method, providing a partly independent confirmation of both results.
\end{abstract}

\maketitle

\thispagestyle{fancy}


%


\begin{figure}
\includegraphics[width=65mm]{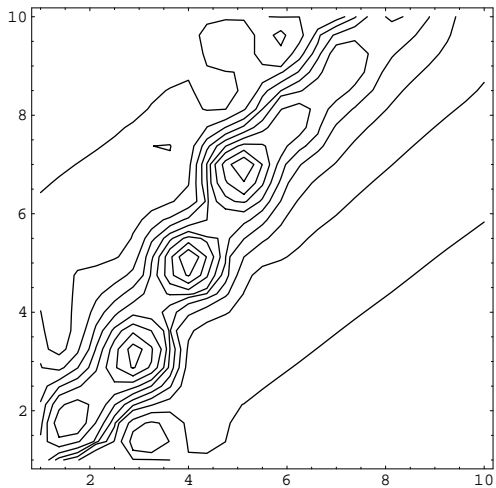}%
\caption{Ridgeline sample with false islands caused by grid too coarse}
\end{figure}
\begin{figure}
\includegraphics[width=65mm]{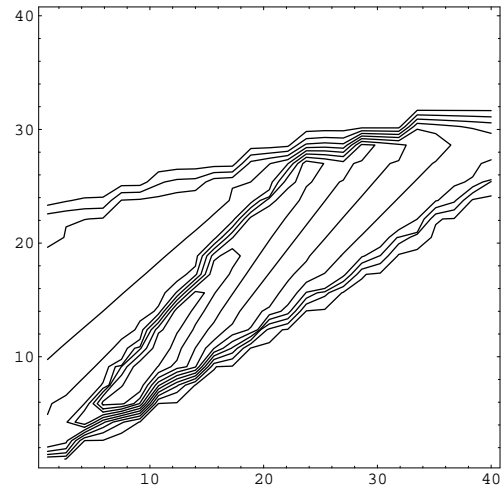}%
\caption{Ridgeline sample near parameter maximum.  Note ridgeline pinching out on right side.}
\end{figure}
\begin{figure}
\includegraphics[width=65mm]{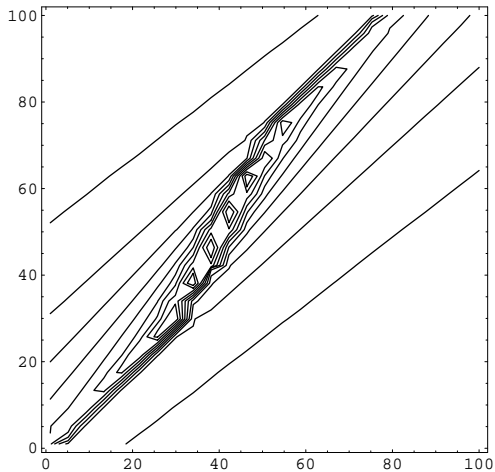}%
\caption{Ridgeline sample showing false islands and intersecting ridgelines.}
\end{figure}
\begin{figure}
\includegraphics[width=65mm]{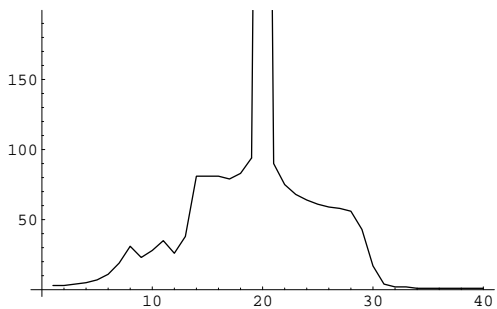}%
\caption{Slice through ridgeline showing narrowness of ridgeline and abrupt cliffs due to fold catastrophe.}
\end{figure}

\section{Introduction} 

In this paper we display and describe two major issues, "ridgelines" and "catastrophes", that arose in our study of the type of extreme mass-ratio binary black hole inspiral signals that are expected to be observed by the gravitational wave satellite observatory LISA when it flies.  These extreme mass-ratio inspiral (EMRI) signals are expected to be among the most scientifically fruitful observations of the LISA mission, making it possible to map out the geometry of the spacetime near a black hole in considerable detail, and also to perform the Ryan\cite{Ryan95} test, which is a go/no-go test of the accuracy of general relativity, with considerable accuracy.  We are carrying out this study by means of a semi-automated grid search, using simulations of the phase and frequency evolution of the EMRI signal.  We compare a target signal to the signal predicted for each combination of possible source parameters in the grid by simulating the entire inspiral.  For each combination, if the phase of a predicted signal differs from the phase of the target signal by less than a specified fraction of a cycle throughout the entire simulated inspiral, that combination is accepted.  Otherwise, the combination is rejected. The shape of these sets of accepted parameters is much more complex than the multidimensional ellipses that are the graphical representation of the results of a Fisher matrix calculation. This paper displays several examples of slices through these multidimensional configurations of all accepted parameters.  Due to space restrictions, we can show only a subset of the larger set of examples displayed at the poster session of this Texas at Stanford conference. An extended paper with more figures is available online at gr-qc/0503063. These figures show a representative sample of the ridgelines and catastrophes that we discovered in our simulations.  As we explain below, these ridgelines and catastrophes are consistent with the nature of the gravitational wave signals and could have been anticipated.  These ridgelines and catastrophes complicate the analysis of EMRI gravitational wave signals and give more detailed results than simple Fisher matrix calculations, but the overall results concerning the accuracy with which parameters can be extracted are in general agreement with the results obtained by Fisher matrix methods.  In particular, the grid search method sometimes finds possible solutions not found by the Fisher matrix method, and the shape of the space of acceptable solutions is almost always different, but the dimensions of the acceptable solution spaces are generally comparable.

In the sections below, we first describe EMRIs and the Ryan test.  We next describe the matched filtering and grid search techniques that will probably be used for analyzing gravitational wave signals. Then we consider why ridgelines, catastrophes and multiple solutions are reasonable expectations in this type of analysis. We also consider various spurious sources of similar signals.  Finally, we briefly consider the implications of these results for using LISA signals to extract source parameters and perform the Ryan test.

\section{Extreme Mass Ratio Binary Inspirals}

LISA is expected to observe stellar mass black holes (7-15 solar masses) spiralling into galactic center black holes of \begin{math}10^6\end{math}to \begin{math}10^7\end{math} solar masses.  These are called extreme mass ratio inspirals ( EMRIs).  Because of the extreme mass ratio, the spin of the smaller black hole is almost negligible.  It has less effect on the phase and frequency evolution of the inspiral than does the quadrupole moment of the larger black hole, which must be measured to perform the Ryan black hole uniqueness test.  To a large extent, the smaller black hole can be treated as a mathematical test particle, and used to indirectly map out the spacetime surrounding the larger black hole.  The most sensitive measurable parameter is the rate of evolution of the phase and frequency of the inspiralling black hole.  As the binary radiates energy in the form of gravitational waves, the two black holes move closer together and the orbital speed increases.  The frequency of the gravitational waves increases synchronously with the orbital frequency, and the rate of radiation increases as the two black holes move closer together and orbit more rapidly.  This results in an accelerating rate of frequency increase, until the two black holes merge.  This accelerating rising frequency signal is called a chirp.

To a first approximation, all chirps look like nearly sinusoidal signals. However, if the orbit is eccentric or inclined with respect to the axis of the spin of the larger black hole, a second fundamental frequency enters the picture, and if the orbit is both eccentric and inclined, a third fundamental frequency is involved.  Complex harmonics and beats of these varied fundamental frequencies are generated, affecting the strength of the gravitational waves and the speed of the frequency evolution. Importantly, the first fundamental frequency can still be detected, and its evolution monitored.  Nevertheless the detailed shape of the waveforms and the details of the phase and frequency evolution depend on a large number of parameters, including the astronomical location of the source on the sky, the location and orientation of LISA, the masses and spins of the two black holes, the magnitude and orientation of each of the two spins, the orientation of the orbit and its periastron, and the timing of the system.  The largeness of this number of parameters, and the accuracy with which any given parameter must be known to predict as many as a hundred thousand orbits, results in a huge parameter space that is much too large for an exhaustive search.

\section{The Ryan Test}

According to the black hole uniqueness theorem, popularly expressed as "a black hole has no hair", in general relativity, a black hole is uniquely determined by its mass, its charge, and its spin or angular momentum, but astrophysical black holes are expected to have negligible charge.  Thus if you know a black hole's mass and spin you can predict its quadrupole moment (and all higher multipole moments as well).  If you can measure a black holes mass, spin and quadrupole moment, you can test this theorem, and hence test general relativity. Ryan\cite{Ryan95} showed mathematically how this can be done by measuring the first few derivatives of the frequency evolution of a chirping EMRI black hole binary.  

\section{Signal Detection and Parameter Extraction by Matched Filtering}

It is expected that the signal-to-noise ratio for LISA EMRIs will be so small that the signal will not be obvious.  Instead, it will be necessary to extract the signals by the matched filtering technique.  For our purposes, it will suffice to consider only signals of the chirp form, i.e. quasi sinusoidal signals with a slowly increasing frequency.  For these signals, the power detected is of order 90 percent of the maximum possible if the signal and the filter are both sinusoidal and they remain in phase by less than 5 percent of a cycle.  On the other hand, the detected signal power drops to zero if the two signals are in quadrature or out of phase by a quarter cycle.  Therefore, we assume that a phase difference of .05 is not detectable, but that one of .25 is clearly detectable.

\section{Multidimensional Grid Search Techniques}

	\subsection{Exhaustive Search Impractical}  

As indicated above, the parameter space of EMRI signal analysis is multidimensional.  As many as 17 different parameters may be involved, but some are negligible.  The effective dimension has been estimated at 14.  The total number of possibilities has been estimated to be as high as \begin{math}10^{40}\end{math}\cite{jon04}.  Even if it is only \begin{math}10^{30}\end{math}, exhaustive search is clearly impractical.  Alternate techniques, probably including some form of hierarchical search, are clearly necessary.  The technique of ignoring overtones and beats and searching for a single fundamental has already been mentioned.  A technique of LISA analysis based on matched filtering of three-week segments with simple polynomial filters followed by a search for connected segments has also been proposed. \cite{WenGair}\cite{jon04}

Even after a signal has been found, extracting the most likely source parameters, which is part of the problem considered in this paper, still requires searching a huge remaining parameter space.  Finding all acceptable parameters, or finding the range of acceptable parameters, requires an even more thorough search.  

	\subsection{Mesh Refinement}

One obvious technique, which has been used effectively in many other contexts, is mesh refinement.  We use this technique extensively, and the major drawback we have found is that, since the solutions tend to fall on very narrow ridgelines,  it is easy to totally miss a solution if your initial grid is too coarse.  Of course, the finer the grid, the greater the computation cost. This has proven to be a serious limitation on the use of this technique in two ways:  it makes us begin with much finer grids, and it makes it difficult to automate the location and sizing of subsequent grid refinements.  Nevertheless, manual grid refinement and relocation is the fundamental technique we used for the searches in this paper.

	\subsection{Hierarchical Search by Dimensional Reduction}

In this technique, one searches in only a few dimensions at a time and then switches to at least partly orthogonal dimensions.  Searching in only one dimension at a time, it is easy to get stuck at any one point on a diagonal ridgeline, when the true maximum is located a diagonal distance away on the same ridgeline.  Searching at least two dimensions at a time avoids this trivial trap and has worked fairly well in the searches we have tried.  
Searching only a single step in a hypercube surrounding the current position is another possibility we have considered, but not yet implemented.  We are not aware of any foolproof technique for following these ridgelines to their true maximum.  The display of false islands resulting from too coarse a grid indicates how easy it is to get stuck at a false maximum.  This has prevented us from implementing an effective fully automated search strategy and forced us to use very fine grids to make sure we were not stuck at a false maximum.

\section{Characteristics of EMRI signals.}

	\subsection{Characteristics of Actual Signals}

The actual signals are still only known to a certain level of approximation; however, the known level is more than enough for our purposes.  In fact as indicated, we truncate the known results at an early post-Newtonian order in order to conserve computation time at an acceptable cost in accuracy.  It is also clear that the actual signals are monotone increases in frequency and also in the first, and probably second, time derivative.  However, the relative effects of spin and eccentricity are not linear, as spin in particular tends to have dominant quadratic terms.  The full expressions of the current best known analytic approximations have many more higher order terms than we have used.  Even the early post-Newtonian order terms that we keep have coefficients that are qudratic and cubic and beyond in spin, eccentricity and the two masses. The relevance of this to our simulations is simply the fact that higher order equations have multiple solutions.  Often only one solution will be physically reasonable, but sometimes multiple solutions yield acceptable results. In our experience, this happens more often with noisy data and with short duration data.  

	\subsection{Characteristics of Simulated Signals}

As indicated, most of the simulations in this paper are based on formulae truncated at fairly early orders.  Since the lower order terms are dominant except in the late stages, low-order polynomial fits are quite good, especially for short intervals in early stages of the inspiral.  Thus it is reasonable to understand some of the behaviors of the simulations in terms of low-order polynomials, even though the actual simulations involve higher order terms and even non-polynomial terms.  This makes it easier to explain the appearance of catastrophes and multiple solutions.  The ridgelines also follow from approximate dimensional analysis and the hierarchical convergence of the various terms, which makes the Taylor expansion effective. 

\section{Simulation Techniques}

We simulate only the frequency evolution of the EMRI phase signal, ignoring all waveshape information.  We use a standard modified Taylor expansion that has been developed by many authors over the years and has been computed by both the post-Newtonian technique and the perturbative technique.  We are indebted to Finn and Thorne\cite{FinnThorne}, Poisson\cite{Poisson97}\cite{Poisson96}, Mino et. al.\cite{JS}, Barack and Cutler\cite{BarCut}, and Glampedakis, Hughes and Kennefick\cite{GHK}\cite{GK} for eccentric evolutions.  
We carry terms up to one order beyond the first appearance of, or to the second appearance of, the spin, the eccentricity, and the quadrupole moment.  We have experimented previously with carrying higher order terms, which give greater absolute accuracy but have much less effect on the relative phase that determines our results.  We now truncate these higher-order terms to reduce computation time.

\section{Factors affecting accuracy and computation costs}

Factors that affect accuracy and computation costs include the order of terms in Taylor series expansions, the use of non-polynomial terms such as logarithms and exponentials, the order of polynomial interpolations, the accuracy of numerical integrations, and the general accuracy of the Mathematica\cite{MM} software package that we used. We tried many combinations of adjustments to these factors. We believe that an adequately accurate, but also reasonably economical combination of these factors was used to produce these graphs.

\section{Notes on Graphs}

Most of the simulations for this paper explored the four-dimensional parameter space spanned by the two masses, the spin, and the eccentricity, or equivalently, the chirp mass, the mass ratio, the spin, and the eccentricity.  (Other parameters that we have explored in other simulations are start and stop times and the quadrupole moment.) Most of the graphs in this paper show the two-dimensional slice in the spin eccentricity plane for a fixed chirp mass and a fixed mass ratio. 
The vertical dimension, which transforms to the inside of the contours on the contour charts, is basically a time dimension.  (It is actually number of cycles, which is monotonically, but not linearly, related to time.)  As long as the inspiral signal generated from a given set of parameters remains inside the acceptable zone, that set of parameters is a possible source for the target signal.  Parameters inside the innermost contour on the contour maps were acceptable for the entire length of time simulated, which corresponds to approximately one year of LISA data.

\section{The fold catastrophe}
The catastrophe we see is the simplest, lowest order catastrophe, called the fold catastrophe.  It can occur anytime a cubic equation or any equation with the S-shape of a cubic is involved.

\section{What We Saw When We Simulated.}

	\subsection{Ridgelines}

Ridgelines are long, narrow, often slightly curving areas inside the acceptable zone.  In a two-dimensional contour plot, they actually look like ridgelines in a topographic map.  In three dimensions they are more like a two-dimensional "fundamental plane", as astronomers use that term; in four dimensions they are three-dimensional, etc. In general, they are one dimension less than the number of dimensions being investigated, which shows that they are the solution to a single equation.  There is one equation for each order in the Taylor expansion, and also one equation for the frequency and each of its derivatives.  In each case the lowest order equation is the most dominant and so it should define the "ridgeline."  In lowest order in the two dimensions defined by the two black hole masses, the ridgeline is the equation for the chirp mass, which can be determined from the frequency and its first derivative.

	\subsection{Catastrophes}

Catastrophes are defined mathematically as the occurrence of a discontinuity in a basically continuous system.  The hysteresis curve is a familiar example.  We see the catastrophes as the sudden jumps in the number of cycles (or length of time) that the signal generated by a set of parameters remains in the acceptable zone as the parameters vary.  Thus the topographic analogy is that the graph looks like box canyons, plateaus and mesas separated by cliffs, rather than hills, mountains and valleys connected by slopes.  Mathematically, catastrophes are usually generated when a solution jumps from one branch of a curve to another or one maximum to another, and that is exactly what is causing the cliffs and plateaus we see in these graphs.  (In the contour maps, the cliffs are multiple contours crowded together.)

	\subsection{Multiple solutions}

Less commonly than the ridgelines and the catastrophes, which are nearly ubiquitous, two or more disconnected islands of high acceptability, or high probability of being a correct solution, will occur.  These seem to occur more often at shorter durations and with higher noise ratios.  Caution is necessary, because too coarse a grid will make a connected ridgeline appear to break up into a series of disconnected islands.  Multiple solutions in mathematics usually result from nonlinear, higher order equations, and many such equations are used to solve the gravitational wave generation formulas and extract the parameters we are interested in.   It is perhaps surprising that multiple solutions do not occur more often.

	\subsection{Intersecting Ridgelines}

Several graphs clearly show a narrower ridge of even higher probability or longer conformance intersecting the main ridgeline.  It is clear that the points inside these boundaries satisfy two separate sub conditions for optimality.  (See figures 1,2 and 3)

\section{Why we could have expected to see these results.}

	\subsection{Ridgelines}

Ridgelines result from the dimensional reduction of the solution space due to the constraints imposed by the equations involved.
There is one equation for each order in the Taylor expansion of the gravitational wave prediction formula.  (These are theoretical quantities.)  There is also one equation for the frequency of the gravitational wave signal and each of its derivatives.  (These are observable quantities.)  By matching these two sets of equations, we can extract the theoretically interesting parameters, such as the two masses, the spin, the eccentricity, and the quadrupole moment from the observed values and hence perform the Ryan\cite{Ryan95} test.
Each equation in a set of multiple equations with multiple unknowns reduces the dimension of the remaining solution space by one. 
An equation with an error bar reduces one dimension to a narrow band instead of to a complete mathematical zero dimensional collapse.  If one equation is a much stronger constraint than the others, one dimension will be reduced by much more than the others, giving the patterns we see.  Since we expect a hierarchy in the strength of the constraints, we should expect a hierarchy in the widths of the dimensions.  At least the first two layers of this hierarchy are visible in figures 1,2 and 3. 

	\subsection{Catastrophes and Multiple Solutions:}  

These are logical possibilities as soon as you have equations of degree two or more for multiple solutions and three or more for catastrophes.  Perhaps it is surprising that catastrophes seem to be so much more common than multiple solutions.  The extent to which the chirps and their derivatives are monotone may promote the tendancy for unique solutions. The fact that spin and eccentricity have many of the same effects on phase evolution (and hence can substitute for each other), but have a different dominant power contributing may allow for the non-monotone type of equation that makes catastrophes possible, particularly in the spin-eccentricity plane.

\section{Implications for Ryan Test}

The fact that the size of the acceptable zone decreases as duration increases, (see narrow peak at the top of figure 4), suggests that EMRI inspirals that are observed for a year or more will give strong results for the Ryan test.

\section{Implications for Parameter Extraction}

The same narrow peak augurs well for LISA parameter extraction as well.
On the other hand, the broad base will be relevant for all short duration observations, including LIGO and other ground based gravitational wave observatories.  

\section{Implications for Hierarchical Search}

The three-week observations proposed as part of the hierarchical search strategy clearly fall in this broad base if treated as individual, disconnected observations.
However, there does not seem to be any reason to continue to analyze them in isolation if a connected signal is found.

\section{Implications for LISA Weak Signals and Noise Analysis}

Some cases of marginally observed LISA signals will fall into this broad base.  This will limit the accuracy with which these signals can be retrodicted as well as predicted, and will hence somewhat hamper the ability to reduce the LISA noise level by subtracting out these signals.  

\section{Conclusion}
 
The ridgelines and catastrophes illustrated in this poster paper show that the acceptable solution space is much more complex than can be represented by a multidimensional ellipsoid.  This complexity will have to be considered during parameter extraction when gravitational wave signals are observed.

%




\bigskip 

\end{document}